\documentclass[preprint,review,12pt]{style}

\usepackage{graphicx}
\usepackage{amsmath}
\usepackage[english]{babel}
\usepackage{lineno}
\usepackage{url}

\setcitestyle{semicolon,aysep={,}}

\begin{document}

\begin{frontmatter}

\title{R{\'e}nyi's spectra of urban form for different modalities of input data}

\author[First]{Mahmoud Saeedimoghaddam}
\ead{saeedimd@mail.uc.edu}

\author[First]{T. F. Stepinski\corref{cor2}}
\ead{stepintz@uc.edu}

\author[First]{Anna Dmowska}
\ead{dmowska83@gmail.com}

\address[First]{Space Informatics Lab, University of Cincinnati, Cincinnati, OH, USA.}

\cortext[cor2]{Corresponding Author}

\begin{abstract}
	Morphologies of urban patterns display multifractal scaling. However, what data should be used to represent an urban pattern and its scaling? Here, we calculated Renyi's generalized dimensions (RGD) spectra using data corresponding to different urban modalities including urban land cover, urban impervious surface, population density, and street intersection points. All data are circa 2010 and we calculated their RGD spectra in six urbanized areas located across the United States. We calculated the RGD spectra by using Hill's numbers rather than statistical moments which leads to a clear interpretation of generalized dimensions and to spatial visualization of pattern's multifractality. The results show that patterns of different urban modalities in a given urbanized area are characterized by different RGD spectra and thus have different morphologies. In our six examples, we found that morphologies of patterns of land cover and impervious surface tend to be monofractal, patterns of street intersection points tend to be moderately multifractal, and patterns of population density tend to be strongly multifractal. Spatial visualization supporting this numerical finding is provided. Thus, when studying the multifractality of urban morphology, it is important to choose a modality that is appropriate to the goal of the investigation. Urban areas may have similar morphologies on the basis of one modality but dissimilar on the basis of another. We have found that two out of our six urban areas have similar morphologies on the basis of all four modalities.  
\end{abstract}


\end{frontmatter}


\section{Introduction}
\label{S:1}

When observing urban agglomerations in high-resolution satellite images it is clear that, while specific details vary, their spatial forms are always heterogeneous and hierarchically self-similar \citep{Haken1995,Batty1989,SAMBROOK2001} with areas of a very low level of urbanization interwoven into areas of a very high level of urbanization. Large number of previous studies showed that urban forms can be described in terms of fractal geometry \citep{Batty1994,Benguigui2000, DeKeersmaecker2003,Thomas2010,Chen2013}.

Closer examination \citep{Appleby1996,Ariza-Villaverde2013} reveals that urban forms are best quantified by a multifractal rather than fractal (monofractal) description. Let assume that the urban system consists of a set of points or pixels each representing a single urban element. If any portion of such set is statistically identical to the original set (statistical self-similarity), the set is fractal, and the urban form has a fractal form. In a fractal all the moments of the probability distribution function scale in the same way so the fractal can be characterized by a single number - a fractal dimension $D$ - which refers to the invariance of the probability distribution function with the change of spatial scale.
However, for the majority of urban systems, the moments of probability distribution function do not scale in the same way and the entire spectrum of numbers -- generalized fractal dimensions -- is needed to characterize the urban form \citep{Hentschel1983}. In such case, the urban form is multifractal. The name multifractal reflects the fact that sparser and denser parts of the urban system have different scaling behaviors.  

We can divide the literature on the multifractal description of urban form on the basis of the type of data used. Different urban forms may be revealed by analyzing different types of data. Possible data modes include classified satellite multispectral images, population grids, street intersections datasets, and night-time lights images. Multispectral satellite images of an urban area can be classified into land cover/land use (LCLU) raster maps as well as used to obtain raster maps of impervious surfaces. LCLU map is converted into a binary map (urban/non-urban pixels); from that binary map, a multifractal spectrum characterizing urban form is calculated \citep{ Murcio2011,Chen2013,Song2019}. A map of urban impervious surface (UIS) is a numerical raster where each pixel has a value between 0 and 1 indicating a share of impervious cover within a pixel. Multifractal characterizations of UIS patterns were performed by \citet{Nie2015} and \citet{Man2019}. Imagery data provides information about the distribution of man-made constructions, consequently, multifractal analysis of imagery data pertains to the spatial form of a built-up area.   

Population data comes from national censuses and is available in the form of aggregated units (shapefile data), however, gridded population datasets, more useful for multifractal analysis, have been derived from original census data \citep{Tatem2017,Dmowska2017b}. Most of the multifractal analyzes of population distribution were conducted for entire countries rather than for urban areas. \citet{Adjali2001} derived multifractal spectra for population distribution for ten different countries. \citet{Mannersalo1998} applied multifractal analysis to the population distribution of Finland, \citet{Semecurbe2016} performed a similar analysis for France, and \citet{VegaOrozco2015} for Switzerland. Only \citet{ChenFeng2017} performed multifractal analysis for the population distribution of a single urban area -- Hangzhou, China. Population data provides information about the distribution of people and thus its multifractal analysis may reveal a different urban structure than the analysis applied to a built-up area.

Recently, street intersection points (SIP) data was used for multifractal analysis of urban form \citep{Murcio2015}. SIP data provides a good proxy for urban form \citep{Arcaute2016,NguyenHuynh2016,Peiravian2017}, especially for studying the long-term temporal evolution of the form, because of the availability of detailed historical maps of road intersections. Such data provides a modality of urban form which is different from those provided by either population data or built-up area data. Finally, \citet{Ozik2005} performed a multifractal analysis of the worldwide distribution of urban areas using night-time lights satellite images. However, the resolution of such data is too coarse to be used for an analysis of a single urban area. 

The contribution of this paper is to compare spatial forms of an urban area as indicated by multifractal analysis of different modes of form-indicating data. Do all data modes indicate the same form or are there significant differences in an urban form depending on what data is used? We have selected six urbanized areas (UAs) throughout the United States. For each UA we calculated the Renyi's generalized dimensions (RGD) spectra of four spatial patterns: land cover (LCLU), percentage of impervious cover (UIS), population density (POP), and street intersection points (SIP). An additional contribution of this paper is a description of RGD spectra in terms of Hill's numbers \citep{Hill1973} instead of statistical moments of probability function characterizing a pattern. As a result, the meaning of generalized dimensions is more transparent. In addition, we can visualize the multifractal structure of spatial patterns as maps of Hill's numbers. 

\begin{figure*}
	\centering
	\includegraphics[width=0.9\textwidth]{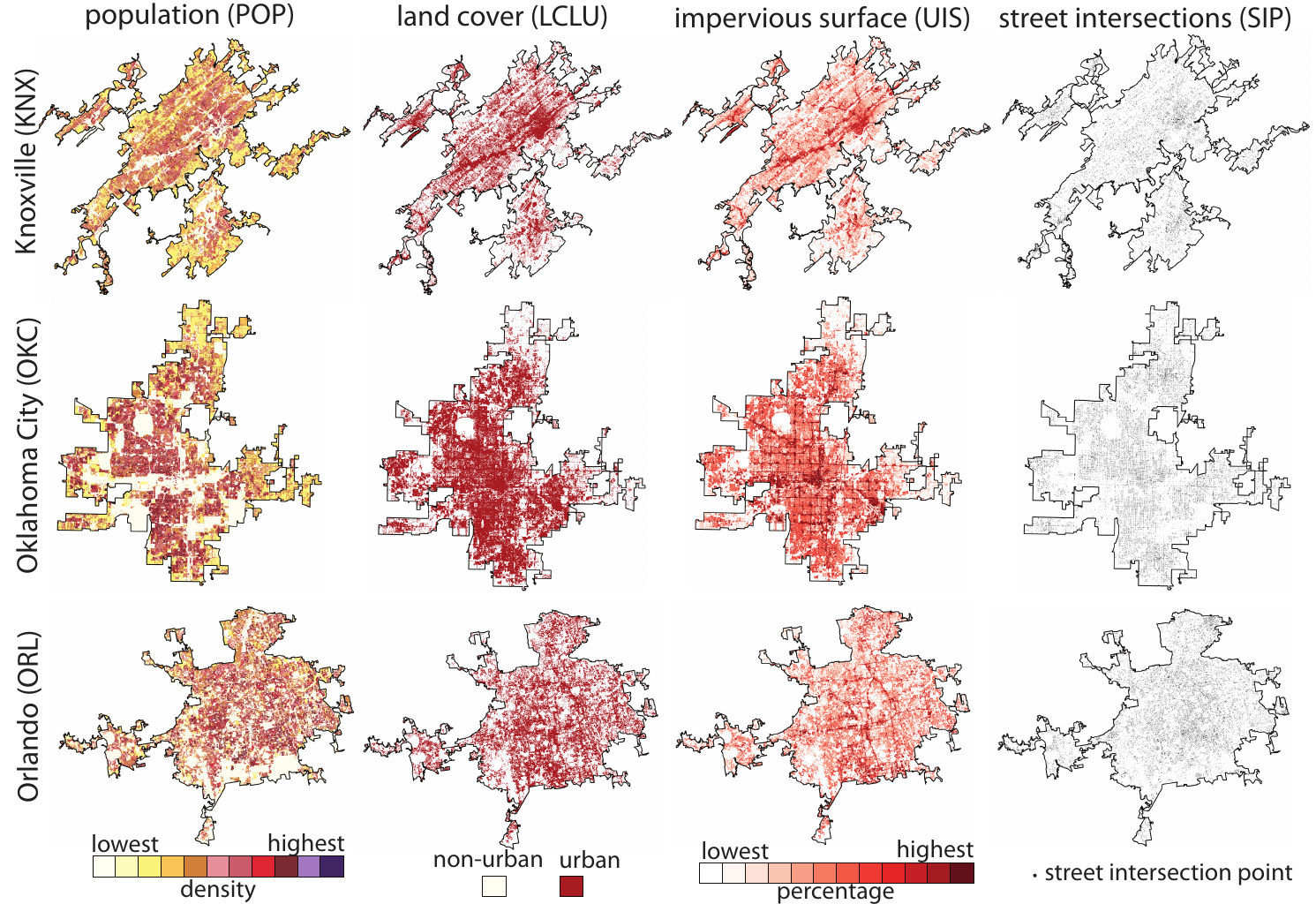}
	\caption{Spatial depiction of four data modes in three out of six urbanized areas analyzed in this study.\label{data}}
\end{figure*}

\section{Data}
We analyze six UAs in their 2010 boundaries, Knoxville, TN (KNX) Oklahoma City, OK (OKC), Orlando, FL (ORL), Philadelphia, PA (PHL), Phoenix, AZ (PHX), and  Portland, OR (POR). The US Census Bureau defines an urbanized area as a densely developed territory with a population over 50,000 that encompass residential, commercial, and other non-residential urban land uses. A UA usually serves as the core of a larger metropolitan statistical area. Within a boundary of each UA, we analyze spatial patterns of LCLU, UIS, POP, and SIP. All data sets are circa 2010.

LCLU and UIS data are from the National Land Cover Database (https://www.mrlc.gov/). We use the 2011 edition of LCLU and UIS as the 2010  edition is not available. 
The LCLU data provides nationwide data on the land cover at a 30m/cell resolution with a 16-class legend based on a modified Anderson Level II classification system. We have reclassified the original LCLU data to only two categories, urban, consisting of three classes (developed low intensity, developed medium intensity, developed high intensity) and non-urban (the remaining 13 classes). In the box-counting method of multifractal analysis, a LCLU ``mass'' of a box is the number of urban cells. The UIS data represent urban impervious surfaces as a share (from 0 to 1) of a developed surface over every 30m cell in the United States. In the box-counting method of multifractal analysis, a UIS ``mass" of a box is the sum of urban impervious surfaces shares for all cells in a box.

The POP data represents population density. It is in the form of 30m/cell grid where each cell has a number of people per cell (the value of the cell could be a fraction). The POP data is available from SocScape (http://sil.uc.edu) and is the result of dasymetric modeling of 2010 US Census Bureau block-level aggregated data \citep{Dmowska2017b,Dmowska2017}. In the box-counting method of multifractal analysis, a POP ``mass" of a box is the sum of all people in cells present in a box (the total number of people in a box). The SIP data starts with 2010 street networks provided by the US Census Bureau (\url{https://www.census.gov/cgi-bin/geo/shapefiles/index.php?year=2010\&layergroup=Roads}). For fixing the topological errors of the street networks we used ArcMap software. ``Line intersections" tool in QGIS software has been employed to extract the junctions (intersections) from the networks. Whereas LCLU, UIS, and POP are grid data, SIP is a point data.  In the box-counting method of multifractal analysis, a SIP ``mass" of a box is the sum of street intersection points in a box.

Fig.~\ref{data} shows patterns of POP, LCLU, UIS, and SIP for three out of six UAs analyzed in the paper. The cells are colored according to respective legends. Whether morphologies of different modalities of urban data are similar or dissimilar and whether the patterns are mono- or multi-fractal is not readily apparent from patterns as shown in Fig.~\ref{data}, but it will become apparent from their RGD spectra.

\section{Methods}
Urban system (for example, an UA) consists of discrete urban elements. In this paper, urban elements are either street intersection points or pixels in maps of land cover, percentage of impervious area, and population density. The spatial pattern of urban elements forms an urban morphology that we want to quantify. We are not interested in an overall layout of an urban area but rather in the statistical properties of its hierarchical structure. For this purpose, we are using multifractal analysis. 

We discuss multifractal analysis through the prism of box-counting method \citep{Cheng1995}. Box counting method is based on a series of grids with different cell (box) sizes; each grid covers the entire pattern of urban elements. For a grid with given box size, ($\epsilon$), a mass (see the previous section) of each box is calculated and stored in the originating box thus transforming a grid to a numerical array. Dividing the mass of each box by the total mass of all urban elements in a region yields a probability distribution $\{p_i\}=\{p_1,\dots p_n\}$, where $n$ is the number of nonempty boxes. Note that both, probability distribution and the value of $n$ depend on the grid size $\epsilon$, dependence that, for a moment, is not reflected in the notation to make the text more lucid. The probability distribution of urban pattern, like all probability distributions, may be characterized by its statistical moments,
\begin{equation}\label{moments}
M_q = \sum_{i}p_i(\epsilon)^q
\end{equation}
\noindent where $\infty \le q \le \infty$ is called a moment order. The multifractal analysis consists of determining whether statistical moments of  $\epsilon$-series of $\{p_i\}$ have power-law scaling with $\epsilon$. If all moments have identical power-low scaling the urban morphology is fractal (monofractal), if they have power-law scaling but with different exponents the urban morphology is multifractal, and if they don't have the power-law scaling the urban morphology is nonfractal. Assuming that the urban structure is either monofractal or multifractal, the results of the multifractal analysis is summarized by the Renyi's generalized dimensions (RGD) spectrum \citep{Renyi1961}.

Most descriptions of the RGD in the urban context \citep{Murcio2015,Salat2017, ChenFeng2017} are in term of $M_q$, however, in our opinion, description of RGD in terms of Hill's numbers \citep{Hill1973} provides more connection to the data and enables spatial visualization of urban structure in terms of RGD. The connection between moments $M_q$ and Hill's numbers $N_q$ is as follows, 
\begin{equation}\label{hillnumber}
N_q =
\begin{cases}
M_q^{1/(1-q)} & \text{if}\ \  q \neq 1 \\
a^{-\sum_{i=1}^n p_i \log p_i}\ \   & \text{if}\ \ q=1
\end{cases}
\end{equation}
\noindent where $a$ is the base of the logarithm used in the equation. 

To better understand the meaning of $N_q$ note that $M_q$ can also be written as $\sum_{i} w_i p_i^{q-1}$, where $w_i=p_i$ but is interpreted here as a weight, so $M_q$ could be interpreted as a weighted mean of the $(q-1)^{\rm th}$ powers of the relative frequencies of urban elements in nonempty boxes and denoted by $\langle p^{q-1} \rangle$. Then, ${\langle p\rangle}_q={\langle p_i^{q-1} \rangle}^{1/(q-1)}$ is the generalized average probability (average relative frequency of urban elements per box). It is called ``generalized" because its value depends on the way the average is defined which, in turn, depends on the value of $q$.

The reciprocal of ${\langle p\rangle}_q$ is the Hill's number $N_q$. Thus, $N_q$ is a number of equally abundant boxes to have average probability the same as the generalized average probability of all boxes. For this reason $N_q$ may be referred to as an {\sl effective} number of boxes. As $q$ decreases from $\infty$ to $-\infty$, ${\langle p\rangle}_q$ decreases from $\max p_i$ to $\min p_i$ and $N_q$ increases from $1/\max p_i$ to $1/\min p_i$. For $q=0$,  ${\langle p\rangle}_q=1/n$ and $N_q=n$. 

$N_q$ is a measure of diversity or disparity of values in the distribution. Small values of $N_q$ indicate distributions restricted to a few high-density boxes, whereas large values of $N_q$ indicate distributions spread over all or the majority of the boxes. For $q \geq 0$, the range of values of $N_q$ is between $n$ (when $q=0$) and 1 (when $q=\infty$ and there is only one box with the maximum value of probability). For $q < 0$, the $\langle p^{q-1} \rangle$ is very small and, consequently $N_q >n$. 

Renyi's generalized entropy $H_q$ \citep{Renyi1961} is related to $N_q$ as follows, 
\begin{equation}\label{Hq}
H_q(\epsilon) =
\begin{cases}
\log N_q (\epsilon) & \text{if}\ \  q \neq 1 \\
{\displaystyle -\sum_{i=1}^n} p_i(\epsilon) \log p_i(\epsilon)  & \text{if}\ \  q=1
\end{cases}
\end{equation}
\noindent where we now explicitly show dependence on spatial scale $\epsilon$. The generalized dimension, $D_q$, is calculated as $D_q = -\lim_{\epsilon\to 0}\ H_q/\log \epsilon$, or, in terms of $N_q$,
\begin{equation}\label{Dqfinal}
D_q =
\begin{cases}
-{\displaystyle \lim_{\epsilon\to 0}} \frac{\log N_q(\epsilon)}{\log\epsilon}  & \text{if}\ \  q \neq 1 \\
\\
-{\displaystyle \lim_{\epsilon\to 0}} \frac{-\sum_{i=1}^n p_i(\epsilon) \log p_i(\epsilon) }{\log\epsilon}  & \text{if}\ \  q=1 
\end{cases}
\end{equation}
Thus, generalized dimensions are the rates of change of effective number of boxes with scale, $N_q(\epsilon) \sim {\epsilon}^{-D_q}$. Because $N_0(\epsilon)= n(\epsilon)$, $D_0$ is a fractal dimension. Other $D_q$ are rates of change of a number of boxes in density-limited subsets of the urban area. The function $D_q(q)$ is referred to as the RGD spectrum; it provides a unique description of an urban pattern.

\begin{figure*}
	\centering
	\includegraphics[width=1.0\textwidth]{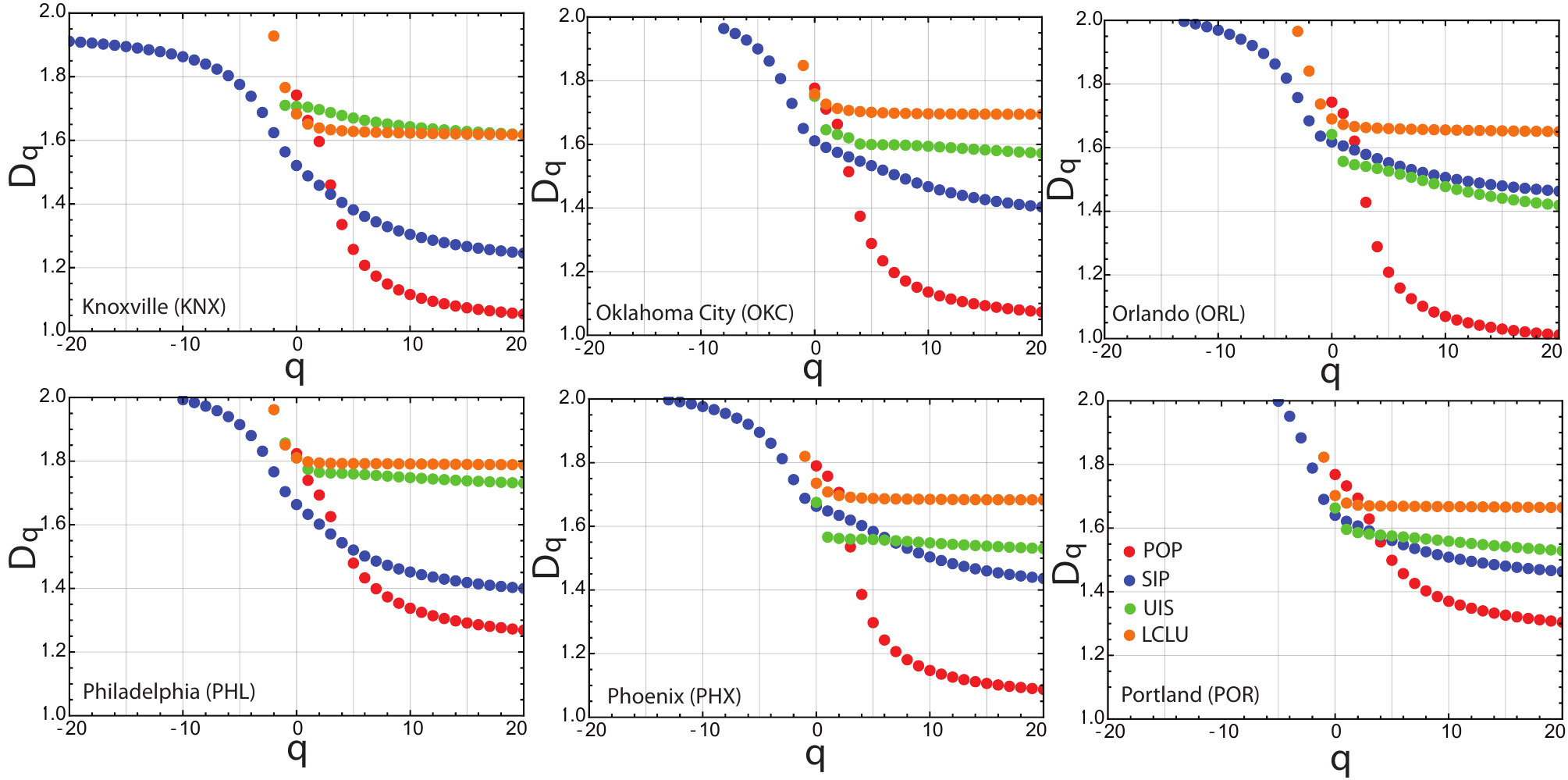}
	\caption{RGD spectra of patterns of LCLU (orange), UIS (green), POP (red) and SIP (blue) in six urban areas analyzed in this study.\label{spectra}}
\end{figure*}

The importance of expressing generalized dimensions in terms of Hill's numbers is that it gives a transparent meaning to generalized dimensions. In addition, for nonnegative orders, the $\epsilon$-series of $N_q$ corresponds to nested subsets $S_q$ of an $\epsilon$-sized grid, such that $S_{\infty}\subset  \dots \subset S_{k+1} \subset S_k \subset \dots \subset S_1 \subset S_0$. The $S_{\infty}$ consists of $N_{\infty}$ densest boxes (could be as little as one box), $S_k$ consists of $N_k$ densest boxes (which include all boxes in $S_{k+1}$), and $S_0$ consists of all nonempty boxes in the $\epsilon$-sized grid. The $\epsilon$-series of maps of these subsets forms a spatial visualization of multifractality of urban pattern. 

\section{Empirical results}
For each UA, we calculated RGD spectra of 2010 SIP pattern, 2010 POP pattern, 2011 UIS pattern, and 2011 urban LCLU pattern. To calculate the RGD spectrum for each dataset, we utilized the globally normalized \citep{Anderson2014} ``enlarged box-counting'' method \citep{Pastor-Satorras1996} with a dyadic sequence of the box sizes $\epsilon$. We calculate values of $D_q$ for $q$ between -20 and 20 in step of 1, but only keep the values fulfilling the following criteria, (1) the resultant value is in the range  ${0 \leq D_q \leq 2}$, (2) $R^2 \geq 0.9$ in the regression estimation of linear fit necessary to translate box counts into an estimate of $D_q$ \citep{Murcio2015}.

Fig.~\ref{spectra} shows the calculated RGD spectra, RGD spectra of SIP patterns are shown in blue, RGD spectra of POP patterns are shown in red, RGD spectra of LCLU are shown in orange, and RGD spectra of UIS patterns are shown in green. The main result is that the multifractality of an urban form depends on the mode of the data used.

\begin{table}[h] 
	\caption{Indices of multifractality $\Delta$ and $\Delta_0$ (slanted font)\label{deltas}}
	\resizebox{8.0cm}{!} {
		\begin{tabular}{c c c c c c c }
			\hline 
			data &  KNX & OKC & ORL & PHL & PHX & POR \tabularnewline
			\hline 
			\hline
			LCLU & 0.3 & 0.15 & 0.32 & 0.17 & 0.14 & 0.16 \tabularnewline
			& {\sl 0.07}    & {\sl 0.06}    &  {\sl 0.04}    &  {\sl 0.02}    & {\sl 0.05}     &  {\sl 0.04} \tabularnewline
			UIS &--  &--  &--  & 0.13 &--  & -- \tabularnewline
			& {\sl 0.09}    &  {\sl 0.18}    &  {\sl 0.22}    &  {\sl 0.08}    &  {\sl 0.14}    & {\sl 0.13} \tabularnewline%
			POP & -- &--  &--  &--  &--  &--  \tabularnewline
			& {\sl 0.69}    &  {\sl 0.70}    &  {\sl 0.73}    &   {\sl 0.55}   &  {\sl 0.70}    &  {\sl 0.46} \tabularnewline
			SIP & 0.67 & 0.56 & 0.53 & 0.59 & 0.56 & 0.53 \tabularnewline
			& {\sl 0.28}    & {\sl 0.21}     & {\sl 0.15}     & {\sl 0.26}     &  {\sl 0.23}    &  {\sl 0.18} \tabularnewline
			\hline
		\end{tabular}
	}
\end{table}

To measure the degree of multifractality we introduce two indices, $\Delta$ and $\Delta_0$. First, $\Delta = \max D_q - \min D_q$, it applies only to patterns where $\max D_q$ corresponds to a negative value of $q$. As we can see from Fig.~\ref{spectra} in many cases the RGD spectrum does not extend to negative values of $q$ due to an insufficient number of data and/or the breakup of the fractal character of the pattern in a sparse portion of an urban area. Recall from the previous section that the right part of the RGD spectrum (between $q=\infty$ and $q=0$) represents an accumulation of the effective number of boxes $N_q$  from those representing the densest part ($q=\infty$) to all boxes ($q=0$). The left part of the RGD spectrum (between $q=-\infty$ and $q=0$) also represents an accumulation of effective number of boxes $N_q$  from those representing the sparsest part ($q=-\infty$) to all boxes ($q=0$).

Thus, if $\max D_q$ corresponds to the negative value of $q$, $\Delta$ represents a difference in scaling exponents of the sparsest and the densest parts of an urban area. However, for patterns where $\max D_q = D_0$ such interpretation of $\Delta$ thus not hold. In such cases we define $\Delta_0 = D_0 - \min D_q$ that represents a difference in scaling exponents of the entire pattern and its densest part. Note that $\Delta_0$ can also be defined for patterns that permit calculations of $D_q$ for negative values of $q$. Table 1 lists values of $\Delta$ and $\Delta_0$ (slanted font). Table 1 reveals that, in all six UAs, RGD spectra for patterns of POP cannot be extended to negative values of $q$. RGD spectra for patterns of UIS cannot be extended to negative values of $q$ in five out of six urban areas.

In all six UAs values of $\Delta_0$ are small for patterns of LCLU. For such patterns values of $\Delta$ are also small with the exception of KNX and ORL. Thus, with possible exception of their sparsest parts, patterns of LCLU are monofractal. This is in agreement with the results of multifractal analysis \citep{Chen2013} of LCLU in Beijing, China. RGD spectra of patterns of UIS are restricted mostly to nonnegative values of $q$ (the single exception is Philadelphia). Small values of $\Delta_0$ indicate that these patterns tent to be only weakly multifractal; the urban area of Orlando is an exception. This is in qualitative agreement with the results of multifractal analyzes \citep{Nie2015,Man2019} for patterns of UIS in Shanghai and Xiamen, China, respectively 

Patterns of POP have RGD spectra restricted to nonnegative values of $q$. However, they are all strongly multifractal as indicated by large values of $\Delta_0$. Previous multifractal analysis of POP for urban area of Hangzhou City, China \citep{ChenFeng2017} also found strong multifractality. Unlike in our study, these authors calculated values of $D_q$ for negative values of $q$ with large absolute values. This resulted in values of $D_q$ as high as 3.5 which is not interpretable in a 2D geometry. Restricting their results to $D_q \le 2$ eliminates values of $D_q$ for negative values of $q$ and makes their result very similar to ours.    

Finally, for patterns of SIP, their RGD spectra extend to negative values of $q$ so values of both, $\Delta$ and $\Delta_0$ can be calculated. Values of $\Delta$ are large indicating strong multifractality. However, the values of $\Delta_0$ are significantly smaller, so there is a large difference in scaling exponents between sparsest and densest parts of the SIP pattern, but a large part of this difference is due to high scaling exponents for the sparsest parts of that pattern. \citet{Murcio2015} performed multifractal analysis of SIP for London as bounded by an external boundary (the green zone) at different years starting from 1786 and ending in 2010. Qualitatively their results are in agreement with ours inasmuch as they indicate strong multifractality for earlier years, and their RGD spectra extend to negative values of $q$. However, their values of $D_q$, for positive values $q$, are, in general, larger than ours indicating that London, even in the past, had more space-filling network of streets than the six present-day American cities analyzed in this paper.

\begin{figure*}
	\centering
	\includegraphics[width=1.0\textwidth]{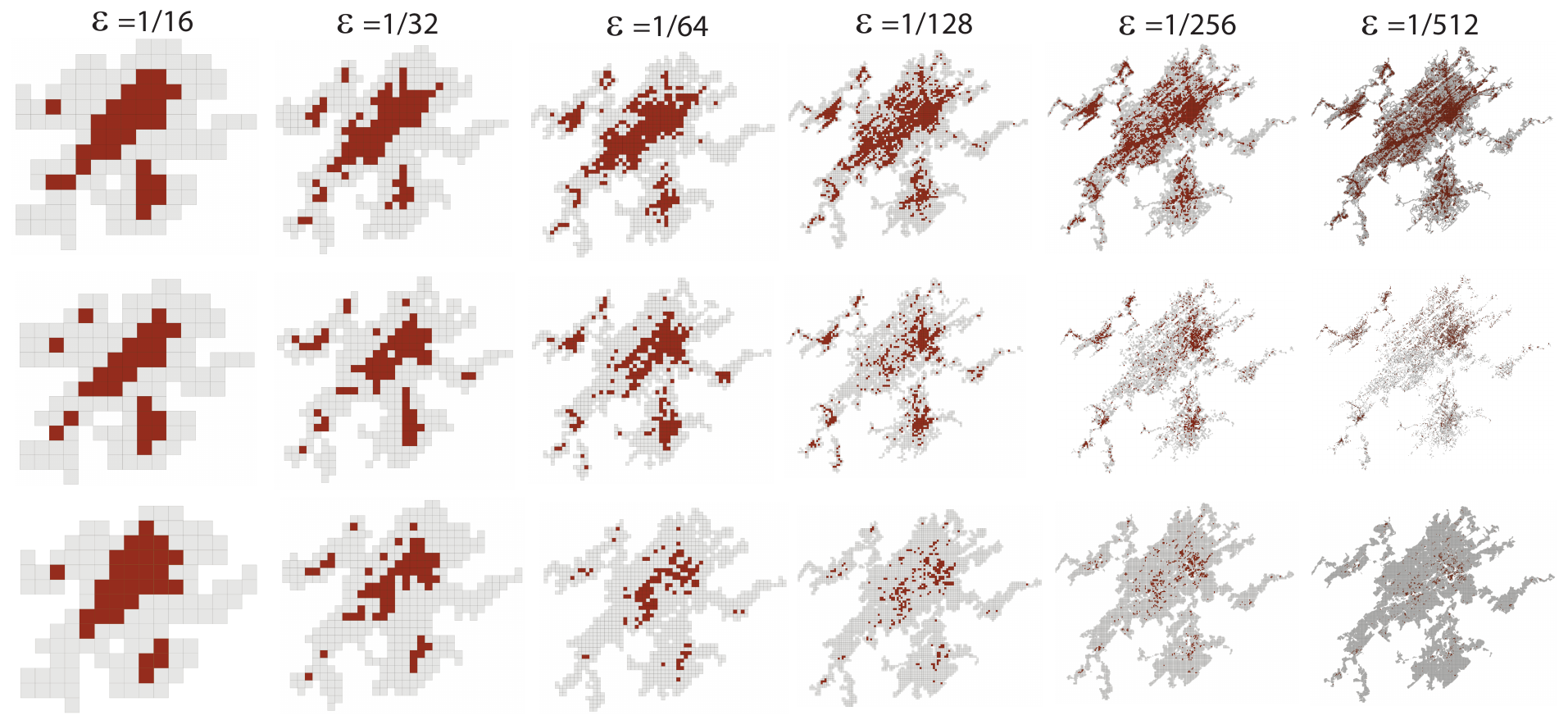}
	\caption{Visualization of multifractality of urban structure in Knoxville, TN using $\epsilon$-series of patterns of boxes; gray color indicates all boxes and brown color indicates $N_{10}$ densest boxes. The top row pertains to UIS, the middle row pertains to SIP, and the bottom row pertains to POP.\label{visual}}
\end{figure*}

\section{Visualizing multifractal structure}
Recall that expressing generalized dimensions in terms of Hill's numbers allows spatial visualization of the urban area's multifractal structure. We demonstrate such visualization for patterns of UIS, SIP, and POP in KNX. Fractal dimensions (values of $D_0$) for these three patterns are 1.71, 1.52, and 1.74, respectively. Values of $D_{10}$, which can be interpreted as a fractal dimension of pattern consisting of only denser part of the urban area, are 1.64, 1.30, and 1.12, respectively. Thus, as the level of space-filling of the entire pattern is the highest for POP closely followed by UIS, and the lowest for SIP, the level of space-filling of the denser part of the pattern is highest for UIS, lower for SIP, and significantly lower for POP. 

Fig.~\ref{visual} depicts three $\epsilon$-series of maps showing effective boxes, $N_q(\epsilon)$, for the entire pattern ($N_0(\epsilon)$ boxes shaded gray) and for the denser part of the pattern ($N_{10}(\epsilon)$ boxes shaded brown). The upper row shows the $\epsilon$-series for UIS, the middle row shows the $\epsilon$-series for SIP, and the bottom row shows the $\epsilon$-series for POP. First, let's focus on the evolution of the gray area (it is also present beneath the brown area) as $\epsilon \rightarrow 1/512$ (the smallest value of $\epsilon$). Depending on the mode of data the gray area evolves differently. For POP and UIS it converges to relatively high space-filling shapes characterized by values of $D_0$ equal to 1.74 and 1.71, respectively. For SIP it converges to a significantly less space-filling shape characterized by $D_0 =1.52$.

Next, let's focus on the evolution of the brown area as $\epsilon \rightarrow 1/512$. Again, the result depends on the mode of the data. For UIS, it converges to a relatively high space-filling shape characterized by $D_{10}=1.64$. For SIP, it converges to a less space-filling shape characterized by $D_{10}=1.30$. For POP, it converges to the least space-filling shape characterized by $D_{10}=1.12$. Fig.~\ref{visual} clearly shows the multifractal structure of all three patterns and it also shows that each pattern has a different multifractal character.

\section{Discussion}
It is now well established \citep{Appleby1996, Hu2012, Ariza-Villaverde2013,Dai2014,Pavon2017} that urban forms displays multifractal scaling. Does this mean that we can compare urban areas by comparing their multifractal spectra? The difficulty is that many previous authors used different data to represent urban form. We have performed multifractal analyzes using four different data modalities (urban land cover, urban impervious surface percentage, population density, and street intersection points) in six urbanized areas across the U.S. Our results shows that using different data modalities results in different multifractal descriptions of urban form. Thus, when using multifractal spectra to characterize and compare different urban areas, the choice of appropriate data is needed. That choice would depend on what aspect of urban character needs to be described and compared.

In many applications, the pattern of population density is of interest. It is important to note that population data used in this paper is the {\sl residential} population data reflecting where people live rather than where they work. We have found that residential population density has strong multifractal scaling; whereas average over six UAs is $\langle D_0^{\rm POP} \rangle=1.77$ (standard deviation $0.03$), $\langle D_{10}^{\rm POP} \rangle=1.20$ (standard deviation $0.13$). This means that densely populated part of UAs is characterized by spatially intermittent pattern while the entire population pattern (regardless of density) is more continuous. This pertains to all six UAs that we have analyzed; the bottom row in Fig.~\ref{visual} shows POP multifractal scaling in the Knoxville UA. 

On the other hand, patterns of urban land cover are mostly monofractal; 
$\langle D_0^{\rm LC} \rangle=1.73$ (standard deviation $0.05$), $\langle D_{10}^{\rm LC} \rangle=1.69$ (standard deviation $0.06$). This means that the part of a UA with a large density of cells classified as ``a developed area" scale similarly to the entire UA. Why the pattern of urban land cover is monofractal whereas the pattern of population density is multifractal? It is the categorical character of land cover data that is responsible for the difference. The class ``developed area" is very general, it does not describe the kind of development (residential, industrial, communal) and it does not recognize between the high and low density of residential development. 

\begin{figure*}
	\centering
	\includegraphics[width=1.0\textwidth]{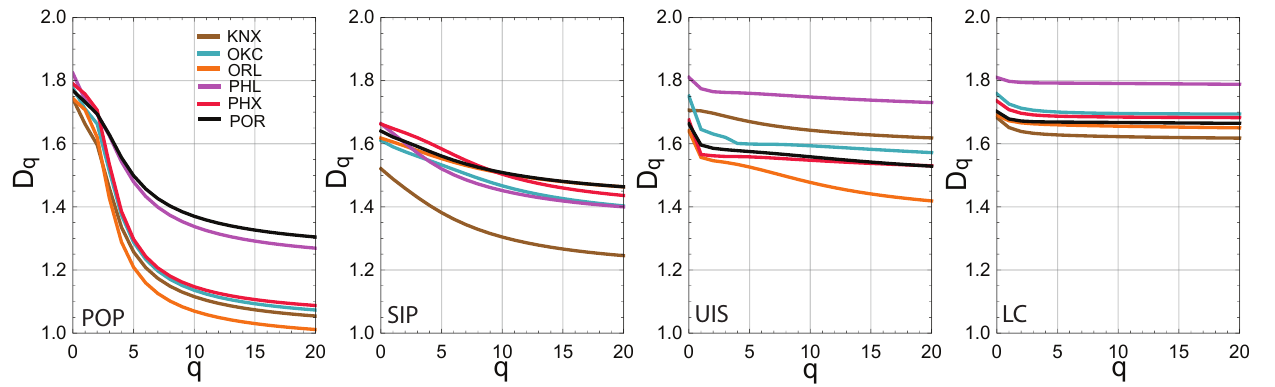}
	\caption{Comparison between six urban areas analyzed in this study based on four data modes. The RGD spectra are restricted to positives values of $q$ for the sake of comparison.\label{compare}}
\end{figure*}

Patterns of urban impervious surface is weakly multifractal; $\langle D_0^{\rm UIS} \rangle=1.71$ (standard deviation $0.06$), $\langle D_{10}^{\rm UIS} \rangle=1.59$ (standard deviation $0.06$). The difference between LC and UIS is that LC is a binary data (urban/non-urban) and UIS is a numerical data (a share of developed space in a cell). As a result, the part of UA consisting of cells with high shares of developed space forms a pattern that is slightly more intermittent than the entire UIS pattern (all cells regardless of their shares of developed space). This is illustrated in the top row in Fig.~\ref{visual} for the Knoxville UA. 

Finally, the patterns of street intersection points are moderately multifractal; $\langle D_0^{\rm SIP} \rangle=1.62$ (standard deviation $0.05$), $\langle D_{10}^{\rm SIP} \rangle=1.46$ (standard deviation $0.08$). SIP is a proxy for the pattern of the street network and thus pertains to the form of communication infrastructure. SIP patterns have smaller fractal dimensions than other patterns in our study. The part of UA with the high density of SIP forms a pattern that is somewhat more intermittent than the entire UIS pattern. This is illustrated in the middle row in Fig.~\ref{visual} for the Knoxville UA.

Why the morphologies of POP patterns are so different from morphologies of LC and UIS patterns? The main reason is that whereas LC and UIS are strictly two-dimensional projections of on-the-ground reality (they are inferred from images), population count indirectly reflects the third dimension -- the hight of residential buildings -- which influences population density. Thus, the variation in values of POP could be much larger than variations in values of LC or UIS. The same argument applies to comparison of morphologies of POP and SIP patterns, however, density of SIP tend to be more correlated with population density than densities of urban LC and UIS are resulting is less difference between morphologies of POP and SIP patterns.  

Structures of urbanized areas can be compared using not one dataset but a collection of RGD spectra corresponding to different data modalities, Fig.~\ref{compare} shows the comparison between six UAs analyzed in this paper based on four data modes. This figure uses the same RGD spectra shown in Fig.~\ref{spectra} but grouped by data mode rather than by UA. We have also restricted the RGD to positives values of $q$ to compare spectra on the same domain of $q$. Recall that small values of $D_q$ correspond to spatially intermediate pattern whereas large values of $D_q$ correspond to the more continuous pattern. Also, recall that values of $q$ decreasing from 20 to 0 represent a cumulatively larger share of UA starting from the densest. 
From Fig.~\ref{compare} we observe that OKC and PHX have similar RGD spectra for all data modalities. It follows that the scaling and hierarchy of corresponding patterns (POP, SIP, UIS, and LC) in these two UAs are all similar\ indicating that these two UAs are structurally similar across multiple domains of interest. On the other hand, KNX and PHL have dissimilar RGD spectra for all data modalities. RGD curves for PHL are always higher than RGD curves for KNX, which means that all patterns in KNX are more intermediate than patterns in PHL regardless of $q$. This points out to fundamentally different structures of these two UAs.

Another contribution of our paper is expressing $D_q$ in terms of Hill's number $N_q$. For $q \ge 0$ values of $N_q$ have tangible spatial interpretation -- they can be used to delineate approximate part of a UA which scales with $\epsilon$ according to $D_q$. This makes possible a spatial visualization of different scaling in different parts of a UA (see Fig.~\ref{visual}), although only for cumulative parts starting with those having the greatest density. Such visualization makes an intuitive understanding of multifractal scaling in spatial patterns possible. Future work will extent such visualization to cumulative parts of the pattern starting from the sparsest part.

\vspace*{2mm}
{\noindent {\bf Acknowledgments.} This work was supported by the University of Cincinnati Space Exploration Institute.}


\begin{thebibliography}{37}
\expandafter\ifx\csname natexlab\endcsname\relax\def\natexlab#1{#1}\fi
\expandafter\ifx\csname url\endcsname\relax
  \def\url#1{\texttt{#1}}\fi
\expandafter\ifx\csname urlprefix\endcsname\relax\def\urlprefix{URL }\fi

\bibitem[{Adjali and Appleby(2001)}]{Adjali2001}
Adjali, I., Appleby, S., 2001. {The Multifractal Structure of the Human
  Population Distribution}. In: Tate, N.~J., Atkinson, P.~M. (Eds.), {Modelling
  Scale in Geographical Information Science}. John Wiley \& Sons Ltd., pp. 69
  -- 85.

\bibitem[{Anderson(2014)}]{Anderson2014}
Anderson, L., 2014. Improved methods for calculating the multifractal spectrum
  for small data sets. Ph.D. thesis, Colorado State University.

\bibitem[{Appleby(1996)}]{Appleby1996}
Appleby, S., Apr. 1996. Multifractal characterization of the distribution
  pattern of the human population. Geographical Analysis 28~(2), 147--160.

\bibitem[{Arcaute et~al.(2016)Arcaute, Molinero, Hatna, Murcio, Vargas-Ruiz,
  Masucci, and Batty}]{Arcaute2016}
Arcaute, E., Molinero, C., Hatna, E., Murcio, R., Vargas-Ruiz, C., Masucci,
  A.~P., Batty, M., 2016. Cities and regions in britain through hierarchical
  percolation. Royal Society Open Science 3~(4), 150691.

\bibitem[{Ariza-Villaverde et~al.(2013)Ariza-Villaverde, Jiménez-Hornero, and
  Ravé}]{Ariza-Villaverde2013}
Ariza-Villaverde, A.~B., Jiménez-Hornero, F.~J., Ravé, E. G.~D., 2013.
  Multifractal analysis of axial maps applied to the study of urban morphology.
  Computers, Environment and Urban Systems 38, 1 -- 10.

\bibitem[{Batty and Longley(1994)}]{Batty1994}
Batty, M., Longley, P., 1994. Fractal Cities: A Geometry of Form and Function.
  Academic Press Professional, Inc., San Diego, CA, USA.

\bibitem[{Batty et~al.(1989)Batty, Longley, and Fotheringham}]{Batty1989}
Batty, M., Longley, P., Fotheringham, S., 1989. Urban growth and form: Scaling,
  fractal geometry, and diffusion-limited aggregation. Environment and Planning
  A: Economy and Space 21~(11), 1447--1472.

\bibitem[{Benguigui et~al.(2000)Benguigui, Czamanski, Marinov, and
  Portugali}]{Benguigui2000}
Benguigui, L., Czamanski, D., Marinov, M., Portugali, Y., 2000. {When and where
  is a city fractal?} {Environment and Planning B} 27(4), 507--519.

\bibitem[{Chen and Feng(2017)}]{ChenFeng2017}
Chen, Y., Feng, J., 2017. {Spatial analysis of cities using Renyi entropy and
  fractal parameters}. {Chaos, Solitons \& Fractals} 105, 279--287.

\bibitem[{Chen and Wang(2013)}]{Chen2013}
Chen, Y., Wang, J., 2013. {Multifractal Characterization of Urban Form and
  Growth: The Case of Beijing}. Environment and Planning B: Planning and Design
  40~(5), 884--904.

\bibitem[{Cheng and Agterberg(1995)}]{Cheng1995}
Cheng, Q., Agterberg, F.~P., Oct 1995. Multifractal modeling and spatial point
  processes. Mathematical Geology 27~(7), 831--845.

\bibitem[{Dai et~al.(2014)Dai, Zhang, Li, and Wu}]{Dai2014}
Dai, M., Zhang, C., Li, L., Wu, W., 2014. {Multifractal and singularity
  analysis of weighted road networks}. {International Journal of Modern Physics
  B} 28(30), 1450215.

\bibitem[{De~Keersmaecker et~al.(2003)De~Keersmaecker, Frankhauser, and
  Thomas}]{DeKeersmaecker2003}
De~Keersmaecker, M.-L., Frankhauser, P., Thomas, I., Oct. 2003. Using fractal
  dimensions for characterizing intra-urban diversity: The example of brussels.
  Geographical Analysis 35~(4), 310--328.

\bibitem[{Dmowska and Stepinski(2017)}]{Dmowska2017b}
Dmowska, A., Stepinski, T.~F., 2017. {A high resolution population grid for the
  conterminous United States: The 2010 edition}. {Computers, Environment and
  Urban Systems} 61, 13--23.

\bibitem[{Dmowska et~al.(2017)Dmowska, Stepinski, and Netzel}]{Dmowska2017}
Dmowska, A., Stepinski, T.~F., Netzel, P., 2017. {Comprehensive framework for
  visualizing and analyzing spatio-temporal dynamics of racial diversity in the
  entire United States}. PLoS ONE 12(3), e0174993.

\bibitem[{Haken and Portugali(1995)}]{Haken1995}
Haken, H., Portugali, J., 1995. {A synergetic approach to the self-organization
  of cities and settlements}. {Environment and Planning B} 22(1), 34--46.

\bibitem[{Hentschel and Procaccia(1983)}]{Hentschel1983}
Hentschel, H.~G., Procaccia, I., 1983. The infinite number of generalized
  dimensions of fractals and strange attractors. {Physica D: Nonlinear
  Phenomena} 8(3), 435--444.

\bibitem[{Hill(1973)}]{Hill1973}
Hill, M.~O., 1973. {Diversity and evenness: a unifying notation and its
  consequences}. Ecology 54, 427–432. 54, 427--432.

\bibitem[{Hu et~al.(2012)Hu, Cheng, Wang, and Xie}]{Hu2012}
Hu, S., Cheng, Q., Wang, L., Xie, S., 2012. {Multifractal characterization of
  urban residential land price in space and time}. {Appl. Geogr.} 34, 161--170.

\bibitem[{Man et~al.(2019)Man, Nie, Z.Li, Li, and Wu}]{Man2019}
Man, W., Nie, Q., Z.Li, Li, H., Wu, X., 2019. {Using fractals and multifractals
  to characterize the spatiotemporal pattern of impervious surfaces in a
  coastal city: Xiamen, China}. {Physica A: Statistical Mechanics and its
  Applications} 520, 44--53.

\bibitem[{Mannersalo et~al.(1998)Mannersalo, Koski, and
  Norros}]{Mannersalo1998}
Mannersalo, P., Koski, A., Norros, I., 1998. {Telecommunication networks and
  multifractal analysis of human population distribution}. Tech. rep., COST257
  Technical Document (98)02.

\bibitem[{Murcio et~al.(2015)Murcio, Masucci, Arcaute, and Batty}]{Murcio2015}
Murcio, R., Masucci, A.~P., Arcaute, E., Batty, M., Dec. 2015. Multifractal to
  monofractal evolution of the london street network. Phys. Rev. E 92~(6),
  062130.

\bibitem[{Murcio and Rodriguez-Romo(2011)}]{Murcio2011}
Murcio, R., Rodriguez-Romo, S., 2011. Modeling large mexican urban metropolitan
  areas by a vicsek szalay approach. Physica A: Statistical Mechanics and its
  Applications 390~(16), 2895 -- 2903.

\bibitem[{{Nguyen Huynh} et~al.(2016){Nguyen Huynh}, {Makarov}, {Fille Legara},
  {Monterola}, and {Chew}}]{NguyenHuynh2016}
{Nguyen Huynh}, H., {Makarov}, E., {Fille Legara}, E., {Monterola}, C., {Chew},
  L.~Y., Apr. 2016. {Spatial Patterns in Urban Systems}. arXiv e-prints.

\bibitem[{Nie and Liu(2015)}]{Nie2015}
Nie, Q., X.-J., Liu, Z., ., 2015. Fractal and multifractal characteristic of
  spatial pattern of urban impervious surfaces. {Earth Science Informatics}
  8(2), 381--392.

\bibitem[{Ozik et~al.(2005)Ozik, Hunt, and Ott}]{Ozik2005}
Ozik, J., Hunt, B.~R., Ott, E., Oct. 2005. Formation of multifractal population
  patterns from reproductive growth and local resettlement. Phys. Rev. E
  72~(4), 046213.

\bibitem[{Pastor-Satorras and Riedi(1996)}]{Pastor-Satorras1996}
Pastor-Satorras, R., Riedi, R.~H., Aug. 1996. Numerical estimates of the
  generalized dimensions of the hénon attractor for negative q. Journal of
  Physics A: Mathematical and General 29~(15), L391--L398.

\bibitem[{Pavon-Dominguez et~al.(2017)Pavon-Dominguez, Ariza-Villaverde,
  Rincon-Casado, deRave, and Jimenez-Hornero}]{Pavon2017}
Pavon-Dominguez, P., Ariza-Villaverde, A.~B., Rincon-Casado, A., deRave, E.~G.,
  Jimenez-Hornero, F.~J., 2017. Fractal and multifractal characterization of
  the scaling geometry of an urban bus-transport network. Computers,
  Environment and Urban Systems, 64, pp.229-238. 64, 229--238.

\bibitem[{Peiravian and Derrible(2017)}]{Peiravian2017}
Peiravian, F., Derrible, S., 2017. Multi-dimensional geometric complexity in
  urban transportation systems. Journal of Transport and Land Use 10~(1).

\bibitem[{Renyi(1961)}]{Renyi1961}
Renyi, A., 1961. On measures of entropy and information. Fourth Berkeley
  Symposium on Mathematical Statistics and Probability. University of
  California Press, Berkeley, Calif., pp. 547--561.

\bibitem[{Salat et~al.(2017)Salat, Murcio, and Arcaute}]{Salat2017}
Salat, H., Murcio, R., Arcaute, E., 2017. Multifractal methodology. Physica A:
  Statistical Mechanics and its Applications 473, 467 -- 487.

\bibitem[{Sambrook and Voss(2001)}]{SAMBROOK2001}
Sambrook, R.~C., Voss, R.~F., 2001. Fractal analysis of us settlement patterns.
  Fractals 09~(03), 241--250.

\bibitem[{Semecurbe et~al.(2016)Semecurbe, Tannier, and Roux}]{Semecurbe2016}
Semecurbe, F., Tannier, C., Roux, S.~G., Jul. 2016. Spatial distribution of
  human population in france: Exploring the modifiable areal unit problem using
  multifractal analysis. Geogr Anal 48~(3), 292--313.

\bibitem[{Song and Yu(2019)}]{Song2019}
Song, Z., Yu, L., 2019. Multifractal features of spatial variation in
  construction land in beijing (1985–2015). Palgrave Communications, 5(1), p.5.
  5(1), 5.

\bibitem[{Tatem(2017)}]{Tatem2017}
Tatem, A.~J., 2017. {WorldPop, open data for spatial demography}. {Scientific
  data} 4(1), 1--4.

\bibitem[{Thomas et~al.(2010)Thomas, Frankhauser, Frenay, and
  Verleysen}]{Thomas2010}
Thomas, I., Frankhauser, P., Frenay, B., Verleysen, M., 2010. {Clustering
  patterns of urban built-up areas with curves of fractal scaling behaviour}.
  {Environment and Planning B} 37(5), 942--954.

\bibitem[{VegaOrozco et~al.(2015)VegaOrozco, Golay, and
  Kanevski}]{VegaOrozco2015}
VegaOrozco, C.~D., Golay, J., Kanevski, M., 2015. {Multifractal portrayal of
  the Swiss population}. {Cybergeo: European Journal of Geography}, article
  714.

\end{thebibliography}

\end{document}